\documentclass[letterpaper]{edm_template}
\usepackage[letterpaper]{geometry}
\usepackage{fixltx2e}
\usepackage{graphicx}
\usepackage{url}
\usepackage{adjustbox}
\usepackage{comment}
\usepackage[utf8]{inputenc}

\begin{document}
\title{Planning Gamification Strategies based on User Characteristics and DM: A Gender-based Case Study} % AIC: how about this shorter title ?
%% \title{Towards Planning Gamification Strategies based on User Characteristics using Data Mining Techniques: A Gender-based Case Study}

\numberofauthors{6} %  in this sample file, there are a *total*
\author{
% 1st. author
\alignauthor
Armando M. Toda\\
       \affaddr{Durham University, United Kingdom}\\
       \affaddr{University of Sao Paulo, Brazil}\\
       \email{armando.toda@usp.br}
% 2nd. author
\alignauthor
Wilk Oliveira\\
       \affaddr{University of Sao Paulo}\\
       \affaddr{Institute of Mathematics and Computer Science}\\
       \affaddr{Sao Carlos, Brazil}
       \email{wilk.oliveira@usp.br}
% 3rd. author
\alignauthor 
Lei Shi\\
       \affaddr{University of Liverpool}\\
       \affaddr{Liverpool, United Kingdom}\\
       \email{lei.shi@liverpool.ac.uk}
\and  % use '\and' if you need 'another row' of author names
% 4th. author
\alignauthor 
Ig Ibert Bittencourt\\
       \affaddr{Federal University of Alagoas}\\
       \affaddr{Maceio - Alagoas}\\
       \email{ig.ibert@gmail.com}
% 5th. author
\alignauthor
Seiji Isotani\\
       \affaddr{University of Sao Paulo}\\
       \affaddr{Institute of Mathematics and Computer Science}\\
       \affaddr{Sao Carlos - Brazil}\\
       \email{sisotani@icmc.usp.br}
% 6th. author
\alignauthor 
Alexandra Cristea\\
       \affaddr{Durham University}\\
       \affaddr{Department of Computer Science}\\
       \affaddr{Durham, United Kingdom}\\
       \email{alexandra.i.cristea@durham.ac.uk}
}

\maketitle

\begin{abstract}
Gamification frameworks can aid in gamification planning for education. Most frameworks, however, do not provide ways to select, relate or recommend how to use game elements, to gamify a certain educational task. Instead, most provide a ``one-size-fits-all'' approach  covering all  learners, without considering different user characteristics, such as gender. Therefore, this work aims to adopt a data-driven approach to provide a set of game element recommendations, based on user preferences, that could be used by teachers and instructors to gamify learning activities. We analysed data from a novel survey of 733 people (male=569 and female=164), collecting information about user preferences regarding game elements. Our results suggest that the most important rules were based on four (out of nineteen) types of game elements: Objectives, Levels, Progress and Choice. From the perspective of user gender, for the female sample, the most interesting rule associated Objectives with Progress, Badges and Information (confidence=0.97), whilst the most interesting rule for the male sample associated also Objectives with Progress, Renovation and Choice (confidence=0.94). These rules and our descriptive analysis provides recommendations on how game elements can be used in educational scenarios.
\end{abstract}

%\IEEEpeerreviewmaketitle

%% Alexandra: 
% (1) Relation to education needs to be clearer (the word 'education' mentioned, starting with the abstract, in the introduction, etc.
%%Armando: 
%Hi professor, I've reread the text again and included more things on the education part (mainly in introduction and conclusions).

% (2) We need to mention the fact that education, in certain areas, such as STEM, is strongly male-dominated; we can cite papers as well as the expression 'the leaking pipeline', and the Athena-SWAN initiative in the UK;
%%Armando:
%I've added a reference for the STEM part and the leaking pipeline in the introduction. I still don't know how to put the Athena-SWAN in the paper 

% (3) we can say the same thing about games - that certain games are male-dominated, whilst others are preferred by females; and thus to justify why it is important to extract what features are responsible for this, and then apply them consistently in adaptive education;
%%Armando:
%

\section{Introduction}
Gamification has been a widely popular phenomenon in the past few years, being used in various domains, including that of education. Gamification is defined as the use of game elements outside their scope (\textit{i.e.}, games or game playing) \cite{Deterding2011,Seaborn2014}. 
%It has recently attracted educators and other specialists in the education domain especially due to its potentials in motivating and engaging students to perform activities that will achieve predefined learning outcomes \cite{Kapp2012,DeSousaBorges2014,Dichev2017a}. 
However, educators often may not be familiar with specific game-related concepts, or know how to use  game elements, or may not have the resources or time necessary  \cite{Marti-Parreno2016,Sanchez-Mena2016,Toda2018f}. 
A solution is to employ conceptual gamification frameworks \cite{Mora2017}. Still, existing frameworks lack resources and explanations on how to use game elements appropriately \cite{Toda2018f}, especially when considering user preferences affected by demographic differences. Understanding users' characteristics, such as gender, may be especially beneficial, \textit{e.g.}, in STEM education, where the well-known problem of 'the leaking STEM pipeline'\footnote{dropout in STEM education}  occurs \cite{VandenHurk2018}. %% Understanding how the educational content needs to be delivered (\textit{e.g.}, via targeted gamification elements) to different genders may be a first step towards having more women study and work in STEM-related subjects.

%An alternative to gamification frameworks is the bottom-up approach provided by the emergent Data-Driven Gamification Design (DDGD), proposing the use of machine learning and/or data mining algorithms to improve the planning of gamification based on real user data \cite{meder2017}. 
In this paper, we apply a data-driven approach to provide insights into the educational domain, via the research question:
 \textit{``How can gender differences in preferences about gamification elements be used to support gamification design?''} %% Alexandra: why are you not directly mentioning gender here? e.g. 'How can/Can the differences in preferences about gamificatin elements for different genders be used/exploited to support gamification design?'
 %% Armando: I've modified the question as you suggested, yours is better
We conducted a very large survey (808 raw answers) allowing respondents to rank  gamification elements. We based these elements on the works of Dignan \cite{Dignan2011} and Toda \textit{et al}. \cite{Toda2018f}, due to (a) the relatively large number and variety of elements, (b) the availability of synonyms used. Next, we used an unsupervised learning algorithm to generate Association Rules to find patterns within the dataset, in order to understand relations among these elements, based on the users' genders. Our main contributions are:
(a) a survey\footnote{\url{https://forms.gle/hFgTT7kCqBKLqiPd8}} for extracting preferences for gamification for education,  applied  to a large, varied number of respondents; 
(b) extracting gamification elements relevant to the different genders, for the educational domain; (c) extracting relations between these elements, relevant to the different genders; (d) insights into users' acceptance of specific game elements, or groups thereof (and their relations). %The contributions are to be used to support gamification planning by teachers and instructors, as well as support further research in the area. 

\section{Related works}
\label{sec:rel}
As there are very few frameworks focusing on gamification in education domains, we discuss: \textit{(i)} existing models related to game elements, \textit{(ii)} gamification studies  on  user characteristics, \textit{(iii)} planning of gamification. %% Alexandra: here it would be good if you had subsections for each (i), (ii), etc. - it's not clear to me (and most probably to a reviewer) which belongs to which below - they are certainly not ordered.

%\subsection{Existing models}

 Yee and Marczewski both proposed models on how to use game elements,using also large data collections \cite{Yee2006,Tondello2016}; however, their focus is different: they collected (a) players' motivations towards online RPGs and (b) generic gamified applications. \cite{Tondello2016} only provides recommendations of elements for behavioural profiles, but not user demographic characteristics, such as gender. 
Yee's model additionally analysed  behaviours from a gender perspective, but only fo online RPG (World of Warcraft) players exclusively. 
%%Armando: I've changed the period because the multiple regression is not a problem although they applied it only in the context of online rpg players in WoW. His final model is only behavioural (from my understanding).
A recent study by Shi and Cristea \cite{Shi2014,Shi2016a} proposed a model and a set of recommendations based on the Self-Determination Theory \cite{Deci1985}. Their  Motivational Gamification Strategies related game elements with each construct of the SDT, \textit{i.e.}, Autonomy, Competence and Relatedness, and implemented them  \cite{Shi2014}, achieving positive results for each construct. Such studies show how motivational theories and gamification constructs can be related, as well support gamification in education. They however do not support the design process of gamification for educators and teachers.

%% \subsection{Understanding users' characteristics and preferences}
Denden \textit{et al}. \cite{Denden2018} conducted an experiment analysing user preferences (N = 120) over eight game elements within a gamified educational system, based on  personality traits (the famous 'Big Five' \cite{john1999big}). According to the authors, only extraversion, openness and conscientiousness affected students' preferences for particular game elements. The authors also stated the importance of this kind of recommendation to designers and instructors when gamifying their learning environments. However, the gamification in education literature lacks studies which relate the acceptance and influence of game elements with users' genders, involving large-scale data \cite{Albuquerque2017}. One recent study  \cite{Pedro2015}  conducted an experimental study aiming at identifying differences between male and female users (N = 70) towards 'gaming the system' behaviours. It was shown that game elements led to male users decreasing their undesired behaviours; moreover, female users felt less competent than male users. Nevertheless, although the results are interesting, the number of students who were analysed is still relatively small, with students  from within a course context - whereas our study has a wider scale and  variety of participants. 

%% \subsection{Gamification design frameworks in education}
Toda \textit{et al}. \cite{Toda2018f} proposes a framework for blended classroom environments using social networks, via a list of recommendations (names of gamified strategies) based on previous studies in gamification in education. They also apply Dignan's game elements classification \cite{Dignan2011}. However, the gamified strategies proposed are solely based on literature. Nevertheless, their positive results show that game elements are suited for educational environments (\textit{e.g}., classroom and digital platforms). As noted, other gamification frameworks focused on specific domains (\textit{e.g.} Computational Thinking \cite{Kotini2015}).  Klock et al \cite{Klock2016}'s framework is usable for adaptive system. Still, Mora et al \cite{Mora2017} note that this framework focuses on the researchers, rather than the stakeholders (teachers and instructors) and presents limited recommendations on game elements usage.
 
Thus, whilst gamification shows potential benefits for educational applications, the gender differences in preference towards specific game elements needed further, large-scale, systematic studies, to better provide support for  Data-Driven Gamification Design, as tackled by our current paper. 

\section{Dataset and Methods}
\label{sec:met}
%% To conduct this study, we developed a survey to identify to what extent a game element was perceived as being relevant in a game. %For the actual survey, we opted to use the term ``game'' instead of ``gamification'', as the use of game elements, as out experience shows that many people are still not familiarized with the term ``gamification'' nor are able to identify (without additional explanations) gamified applications. However, we made sure that our respondents were informed that these game elements are to be used in educational applications or systems. 
Our survey on game elements contains 29 questions. The first part collected demographic information (age, favourite game setting, and gender). The second part asked to what extend certain game elements were relevant to users in the gamified educational system context, through a Likert Scale, from 1 ``I think this element is irrelevant to me'' to 5 ``I think this element is highly relevant to me''. 
The game elements used and their advantages and potential drawbacks are presented in \url{https://tinyurl.com/y44kqvn5
}  based on  \cite{Dignan2011} and used in  \cite{Toda2018f} in an educational domain. 

Additionally to theoretical motivations, we further validated the selected gamification elements with 4 specialists in gamification, who were also teachers, via an interview, verifying the specialists' acceptance of the used elements, concepts, as well as questions cohesion. Finally, a pilot survey with 18 people verified the time spent and the consistency of the questions, before launching the main survey \url{https://goo.gl/forms/d0i5WosBcMVWvQAK2}. We then recruited surveyees through social networks, forums and  digital environments used by people who  play games. %We considered that all surveyees would have participated in education as learners at some point of their lives, so they would all be able to form opinions in terms of what would be useful to them in that context. Thus, their experience with games was considered in this case more important. 

In total, we collected 808 raw answers. Further cleaning removed data from users who: (a) did not answer all questions; (b) claimed not having played any digital games; (c) were of age$<$0 or age$>$90. Then, we analysed our population characteristics based on demographic data. As the normality test showed a non-normal distribution, a Mann-Whitney test \cite{Mann1947} was used to compare males and females.

Finally, we used association rule mining to analyse the relations amongst our data, based on gender. Unsupervised learning was used as we do not have any predefined labels (outputs) and also to understand the relations between the elements (different from clustering which create groups based on all variables of the dataset).%% The resulting rules are formulated in the form of an IF ${X}$ THEN ${Y}$ logical sentence \cite{Manimaran2015,Tan2004}.
The algorithm analyses the items' frequency (support) and renders a level of confidence, ranging from 0 to 1 (where 1 is the maximum confidence). The confidence can also be supported by conviction \cite{Brin1997}, lift and leverage -– both measuring the independence of items. %In this work, we used this algorithm to analyse the items that were scored by users, aiming at discovering relationships among the game elements, specific for a given user gender. 

\section{Results and Discussion}
\label{sec:res}
\subsection{Gender differences}
After filtering, we retrieved 733 valid answers (90.72\%). We  applied Cronbach's $\alpha$ on the second group of questions (as game elements were based on a Likert scale) and achieved an $\alpha = 0.83$ (high reliability factor \cite{Tavakol2011}).  Our sample is varied in terms of age (ranging from 13 to 68), but limited in terms of experience in playing (at least a year: by design and filtering) and country of origin (Brazil; due to convenience sampling). Nevertheless, the sampling size is much larger than the recommended one (733 $>>$ 384; people playing online games  estimated at 700 mio;  confidence level 95\%). 

We further organised our valid answers into two groups: males (N = 569) and females (N = 164), and verified the distribution of the data using the Shapiro-Wilk normality test. The result showed that our data rejected the null hypothesis (p $<$ 0.05), so we adopted non-parametric tests in further analyses. Table \ref{tab:b} summarises the result.

\begin{table}[!ht]
\caption{Relevance of game elements, averaged per gender}
\centering
\begin{adjustbox}{max width=.45\textwidth}
\begin{tabular}{lllll}
\hline
\textbf{}        & \multicolumn{2}{l}{\textbf{Gender (mean)}} & \multicolumn{2}{l}{\textbf{Mann-Whitney}} \\ \hline
\textbf{Element} & \textbf{Female}       & \textbf{Male}      & \textbf{W}       & \textbf{p-value}       \\ \hline
Point            & 3.76                  & 3.68               & 44836            & 0.429                  \\ \hline
Level            & 4.14                  & 4.21               & 48418            & 0.427                  \\ \hline
Cooperation      & 3.62                  & 3.86               & 52306            & 0.013                  \\ \hline
Competition      & 3.26                  & 3.56               & 53016            & 0.006                  \\ \hline
Renovation       & 4.16                  & 3.78               & 35878            & 2.36e-03               \\ \hline
Progress         & 4.24                  & 4.32               & 48856            & 0.312                  \\ \hline
Objective        & 4.41                  & 4.4                & 45902            & 0.791                  \\ \hline
Puzzles          & 4.14                  & 3.91               & 40636            & 0.008                  \\ \hline
Novelty          & 4.05                  & 4.16               & 49530            & 0.197                  \\ \hline
Chances          & 3.68                  & 3.61               & 44901            & 0.447                  \\ \hline
Social Pressure  & 3.43                  & 3.65               & 51142            & 0.05                   \\ \hline
Acknowledgement  & 3.85                  & 3.73               & 44673            & 0.387                  \\ \hline
Data             & 4.05                  & 4.09               & 46675            & 0.994                  \\ \hline
Scarcity         & 3.16                  & 3.42               & 52468            & 0.011                  \\ \hline
Choice           & 4.07                  & 4.23               & 50267            & 0.08                   \\ \hline
Time Pressure    & 3.16                  & 2.97               & 42711            & 0.09                   \\ \hline
Economy          & 3.41                  & 3.42               & 46738            & 0.973                  \\ \hline
Sensation        & 3.62                  & 3.1                & 37094            & 1.17e-02               \\ \hline
Classification   & 3.51                  & 3.72               & 51340            & 0.042                  \\ \hline
\end{tabular}
\end{adjustbox}
\label{tab:b}
\end{table}

Table \ref{tab:b} shows many significant differences (p-value $<$ 0.05). Interestingly, Competition, Cooperation, Social Pressure, Scarcity and Classification were considered slightly more relevant by the males, whilst Renovation, Puzzle and Sensation elements were considered more relevant by females. Time pressure was disliked by males, but not as much by females.

\begin{comment}
\begin{figure}[!ht]
    \centering
    \includegraphics[width=.4\textwidth]{fig_a.png}
    \caption{Visual representation of differences in perception of game elements between genders. Red = Female. Blue = Male.}
    \label{fig:a}
\end{figure}
\end{comment}

\subsection{Descriptive Analysis}
\label{sub:des}
Comparing surveyees, for elements preferred in different proportions, which are  statistically significant  between males and females,  Cooperation  was  more relevant to males (57.3\%); with 41.6\% males selecting highly relevant, vs. 31.1\% females (Table \ref{tab:c}).

\begin{figure}[!ht]
    \centering
    \includegraphics[width=.45\textwidth]{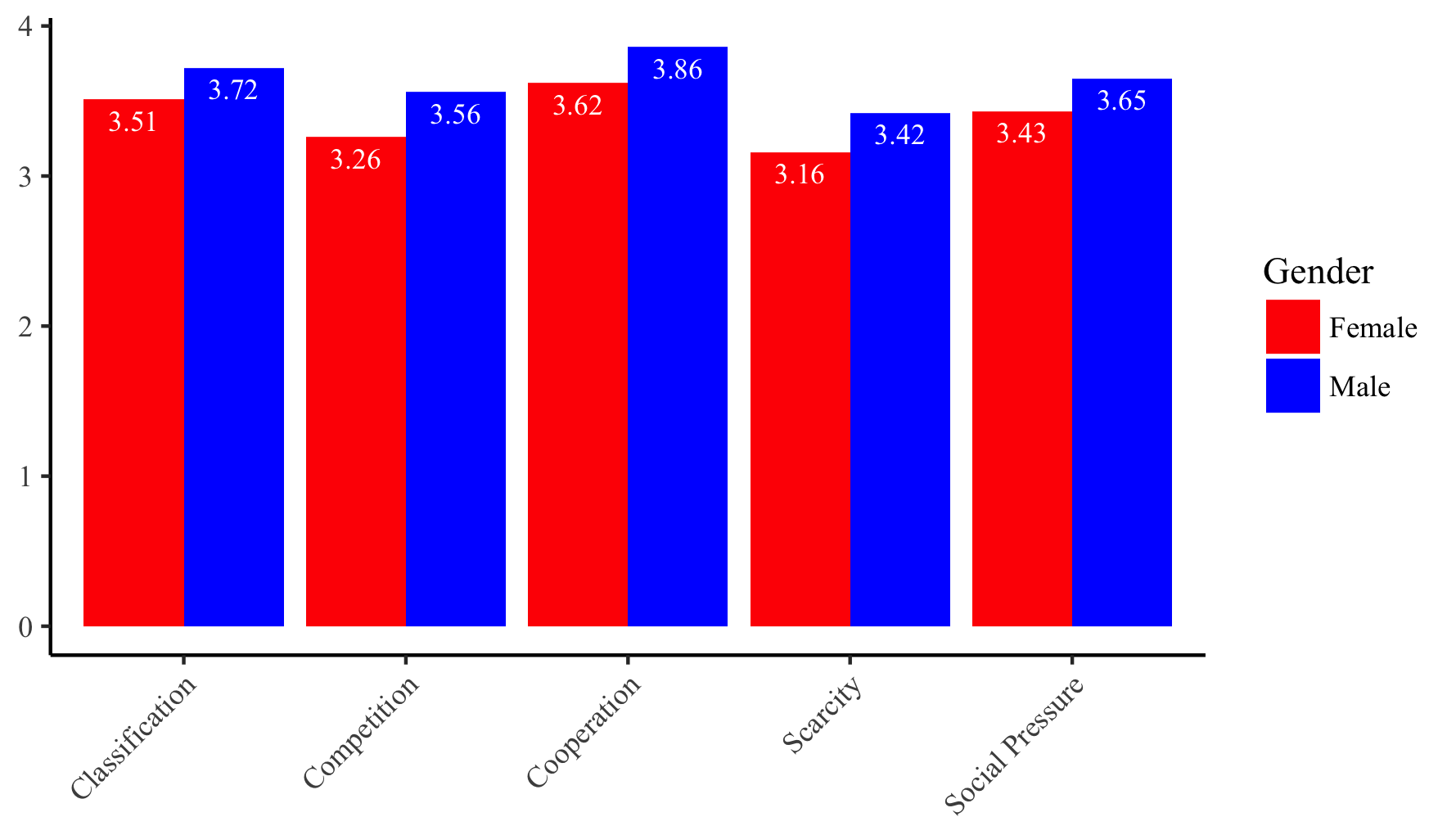}
    \caption{Favourite game elements for males (a,b- correspond to Tables \ref{tab:c},\ref{tab:d}, rsp.) }
    \label{fig:b}
\end{figure}

A more drastic difference appeared when more than 25\% of the females did not consider Competition  to be relevant, versus 54.5\%  males. This result suggests that males may perceive social interactions, such as Competition elements and Cooperation, as highly relevant overall in games, with a slight preference of Competition. For females, however, Competition is not that relevant, but, surprisingly, Cooperation is only marginally relevant. Social Pressure is significantly less liked by females (difference of 7.7\%). %In practice, this information might warn teachers to avoid conducting competitions, if females are present in the classroom, and given them time-pressure-related activities, for instance, instead.

\begin{table}[!ht]
\caption{Cooperation, Competition and Social Pressure answers.}
\centering
\begin{adjustbox}{max width=.5\textwidth}
\begin{tabular}{lllllllllllll}
\hline
\textbf{}      & \multicolumn{4}{l}{\textbf{Cooperation}}            & \multicolumn{4}{l}{\textbf{Competition}}            & \multicolumn{4}{l}{\textbf{Social Pressure}}        \\ \hline
\textbf{Sc} & \textbf{F} & \textbf{\%} & \textbf{M} & \textbf{\%} & \textbf{F} & \textbf{\%} & \textbf{M} & \textbf{\%} & \textbf{F} & \textbf{\%} & \textbf{M} & \textbf{\%} \\ \hline
1              & 12         & 7.3         & 37         & 6.5         & 20         & 12.2        & 42         & 7.4         & 14         & 8.5         & 32         & 5.6         \\ \hline
2              & 20         & 12.2        & 53         & 9.3         & 25         & 15.2        & 80         & 14.1        & 26         & 15.8        & 63         & 11.1        \\ \hline
3              & 38         & 23.8        & 100        & 17.6        & 44         & 26.8        & 137        & 24.1        & 45         & 27.4        & 146        & 25.7        \\ \hline
4              & 43         & 26.2        & 142        & 25          & 43         & 26.2        & 135        & 23.7        & 34         & 20.7        & 160        & 28.1        \\ \hline
5              & 51         & 31.1        & 237        & 41.6        & 32         & 19.5        & 175        & 30.8        & 45         & 27.4        & 168        & 29.5        \\ \hline
\end{tabular}
\end{adjustbox}
\label{tab:c}
\end{table}

Scarcity was instead favoured (significantly) by males (Table \ref{tab:d}), where 37.8\% of the females are indifferent. Classification,  also a social element, was considered significantly more relevant by males. 50.6\% of the females considered it relevant, against 60.6\% males (Table \ref{tab:d}). Based solely on our descriptive analysis, we observed that the male population considered limited or rare tasks, allowing,\textit{ e.g.}, rewards such as  interaction or collecting titles, as relevant. Again, in practice, this information allows the teachers to create titles for completing specific tasks during their lectures \textit{e.g.}, by giving a title of 'Speedster' to the student who completes a list of task correctly and quicker than the others. %However, the teacher would have to allow for equivalent activities that are more appropriate for female students, such as puzzles (see Figure \ref{fig:c}).

\begin{table}[!ht]
\caption{Scarcity and Classification answers}
\centering
\begin{adjustbox}{max width=.45\textwidth}
\begin{tabular}{lllllllll}
\hline
\textbf{}   & \multicolumn{4}{l}{\textbf{Scarcity}}               & \multicolumn{4}{l}{\textbf{Classification}}         \\ \hline
\textbf{Sc} & \textbf{F} & \textbf{\%} & \textbf{M} & \textbf{\%} & \textbf{F} & \textbf{\%} & \textbf{M} & \textbf{\%} \\ \hline
1           & 16         & 9.8         & 32         & 5.6         & 13         & 7.9         & 25         & 4.4         \\ \hline
2           & 25         & 15.2        & 78         & 13.7        & 14         & 8.5         & 68         & 11.9        \\ \hline
3           & 62         & 37.8        & 189        & 33.2        & 54         & 32.9        & 131        & 23          \\ \hline
4           & 39         & 23.8        & 161        & 28.3        & 42         & 25.6        & 163        & 28.6        \\ \hline
5           & 22         & 13.4        & 109        & 19.2        & 41         & 25.0        & 182        & 32.0        \\ \hline
\end{tabular}
\end{adjustbox}
\label{tab:d}
\end{table}

\begin{figure}[!ht]
    \centering
    \includegraphics[width=.45\textwidth]{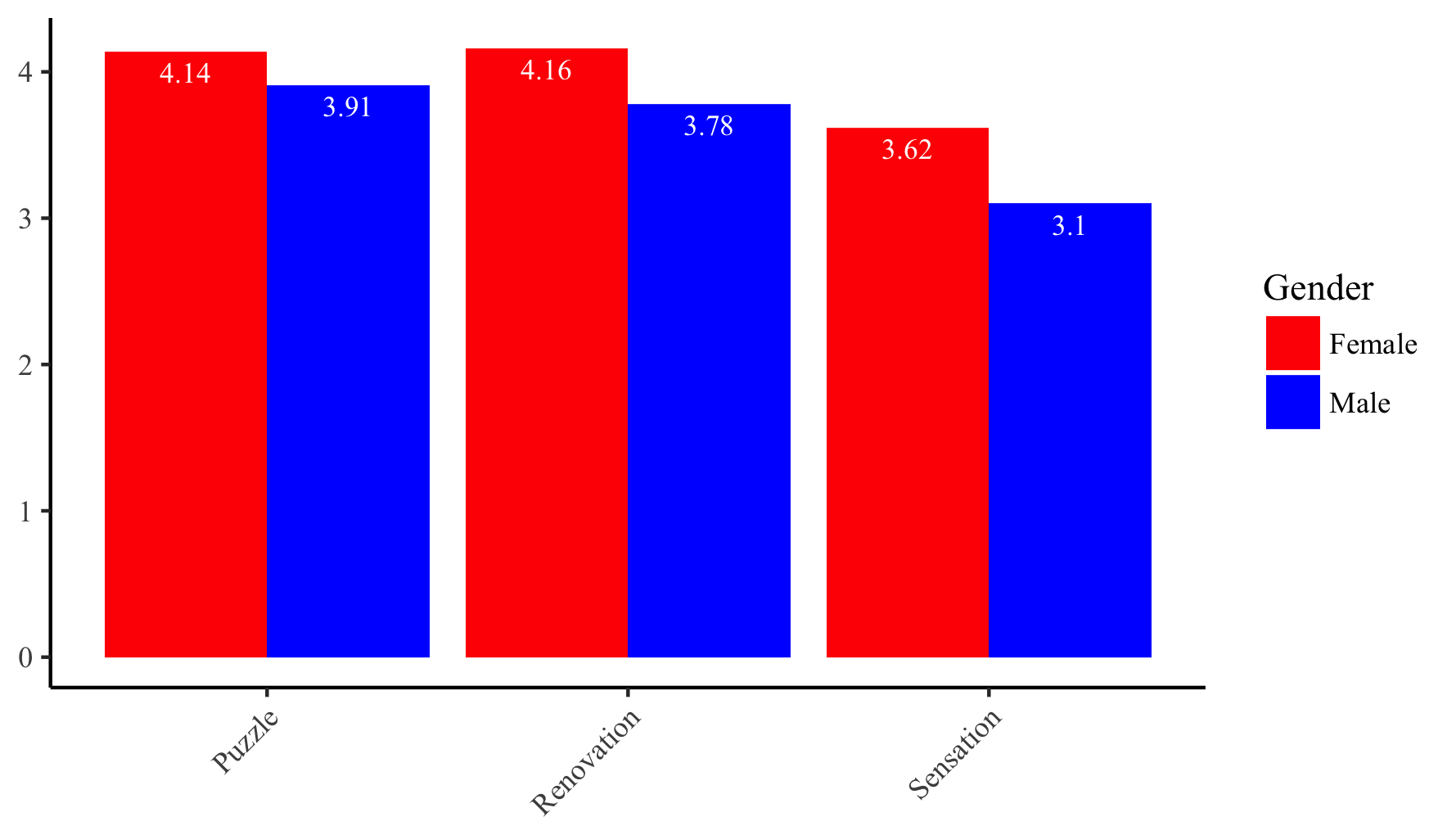}
    \caption{Favourite game elements for the female sample}
    \label{fig:c}
\end{figure}

As for the elements most favoured by females (Figure \ref{fig:c}), Renovation scored highest. 76.2\% said it was relevant, with 50\% considering it highly relevant. In contrast, only 30.0\% of the males considered it highly relevant, with almost 30\% indifferent (Table \ref{tab:e}). %This demonstrates that, in the scope of this study, males cared less about opportunities to redo a task. This can be explicable in the context of our observation, where most of the elements that were found more relevant by males were social; from these, in general, the attraction for males went to ``win-or-lose'' situation, \textit{e.g.}, they cooperate with or compete against someone to win something; they achieve a goal (win situation) and accumulate titles to demonstrate to others their victories. 
%% On the contrary, the Renovation element is more related to the opportunity to improve or master a skill related to a specific activity \cite{Dignan2011}. 
In a learning context, this may tell the teacher that female students might be more pleased with features as ``continue'', ``try again'' or be given 'extra lives'.

Another element  highly relevant to females was Puzzles: 80.8\%, against 68.1\% of males; with 21.9\% males indifferent. Again, females, in this scope, considered that testing their skills was more relevant than males did. The Puzzle and Renovation elements, when combined in practice, allow problem solving, with the opportunity to correct mistakes. 

Finally, the Sensation element was considered more relevant by females. More than half (54.8\%) of the female sample considered it relevant, against 42.5\% of the males. This could be explained by  Sensation  being  related to the user experience  \cite{Dignan2011}, and, based on Table \ref{tab:e}, we can infer that the most relevant elements for the female sample were related to the experience, rather than social ones. This means that they may perceive tasks that involve their senses, \textit{e.g.}, with a visual or phonetic appeal, as more relevant, which could further be redone whenever they wish, to improve a certain skill through challenges. In practice, this means that using materials and resources that are more visual appealing may be more pleasant to female students than the male ones.  

\begin{table}[!ht]
\caption{Renovation, Puzzle and Sensation answers }
\centering
\begin{adjustbox}{max width=.5\textwidth}
\begin{tabular}{lllllllllllll}
\hline
\textbf{}   & \multicolumn{4}{l}{\textbf{Renovation}}             & \multicolumn{4}{l}{\textbf{Puzzle}}                 & \multicolumn{4}{l}{\textbf{Sensation}}              \\ \hline
\textbf{Sc} & \textbf{F} & \textbf{\%} & \textbf{M} & \textbf{\%} & \textbf{F} & \textbf{\%} & \textbf{M} & \textbf{\%} & \textbf{F} & \textbf{\%} & \textbf{M} & \textbf{\%} \\ \hline
1           & 4          & 2.4         & 14         & 2.5         & 5          & 3           & 17         & 3           & 12         & 7.3         & 108        & 19          \\ \hline
2           & 9          & 5.5         & 39         & 6.8         & 8          & 4.9         & 40         & 7           & 21         & 12.8        & 97         & 18          \\ \hline
3           & 26         & 15.8        & 167        & 29.3        & 20         & 12.2        & 125        & 22          & 41         & 25          & 122        & 21.4        \\ \hline
4           & 43         & 26.2        & 161        & 28.3        & 57         & 34.8        & 180        & 31.6        & 34         & 20.7        & 115        & 20.2        \\ \hline
5           & 82         & 50          & 188        & 33          & 74         & 45.2        & 207        & 36.4        & 56         & 34.1        & 127        & 22.3        \\ \hline
\end{tabular}
\end{adjustbox}
\label{tab:e}
\end{table}

\subsection{Association Rules analysis}
%% In this section, we present the rules generated from our dataset, by using the Apriori algorithm. 
To identify the strongest rules for each gender and to verify how the rules found matched or complemented the findings from our descriptive analysis (Section 4.2), we used the Apriori algorithm in Weka, with: \textit{(i)} minimum support of 10\% for the male sample size and 20\% for female (to balance sample sizes); \textit{(ii)} minimum confidence of 90\%; and, after applying those attributes, \textit{(iii)} we used the measures of interest \textit{conviction}, \textit{lift} and \textit{leverage} to find the most interesting rules \cite{Manimaran2015}. Using this setting we found a total of 11 rules for the female sample and 13 rules in the male one. %% We used different minimum support for each subset, due the large variability that would occur in the female sample with only 10\% of minimum confidence (at 10\%, Apriori finds 2112 rules for the female sample).

%\begin{equation}
%    \centering
%    confidence(A \Longrightarrow B) = support(A %\cup B)\div support(A)
%    \label{form:a}
%\end{equation}

The majority ($>$90\%) of the rules were based on the Objective element, which suggests its overall popularity. This translates into a general recommendation towards using 'Objective' elements in the educational gamification design, such as  missions, milestones and quests, to guide the students. In this work we focused on analysing the most interesting rules in female and male samples. For  females, the strongest rules were associated with the Objective (Table \ref{tab:f}). The lift $>$ 1 and leverage near 0 indicate that our items are independent and have a positive correlation, and the conviction between 1 and 5 indicates that these are interesting rules. The strongest rule relates Progress, Acknowledgement and Data elements (\textit{e.g.}, Representations of progression, badges and medals and results screen) with Objective (\textit{e.g.}, missions and quests). Rules regarding Progress and Level were also amongst the 10 strongest.  Thus, we can suggest that teachers and instructors should use Acknowledgement (such as badges and trophies), with other elements associated with the personal enhancement of users (Progress and Level).%% This is conform with findings of non-data-driven evaluations of implementing these elements, which were found to increase motivation for achieving goals \cite{Kapp2012,Toda2018f}. However, with our settings, no rules concerning sensation and renovation were found for the female sample.

\begin{table*}[!ht]
\caption{Relevant association rules for female sample}
\centering
\begin{adjustbox}{max width=\textwidth}
\begin{tabular}{lllllll}
\hline
\textbf{\begin{tabular}[c]{@{}l@{}}Rule \\ ID\end{tabular}} & \multicolumn{1}{c}{\textbf{If}}      & \textbf{Then} & \textbf{Conf} & \textbf{Lift} & \textbf{Lev} & \textbf{Conv} \\ \hline
1                                                           & \{progress, acknowledgement, data\}  & \{objective\} & 0.97                & 1.63          & 0.08              & 7.04                \\
2                                                           & \{level, progress, acknowledgement\} & \{objective\} & 0.97                & 1.62          & 0.08              & 6.84                \\
3                                                           & \{progress, acknowledgement\}        & \{objective\} & 0.96                & 1.6           & 0.1               & 6.04                \\
4                                                           & \{level, acknowledgement\}           & \{objective\} & 0.95                & 1.59          & 0.09              & 5.5                 \\
5                                                           & \{point, acknowledgement\}           & \{objective\} & 0.95                & 1.58          & 0.08              & 4.96                \\
6                                                           & \{progress, puzzles\}                & \{objective\} & 0.94                & 1.57          & 0.1               & 4.73                \\
7                                                           & \{puzzles, novelty\}                 & \{objective\} & 0.92                & 1.54          & 0.07              & 3.82                \\
8                                                           & \{novelty, acknowledgement\}         & \{objective\} & 0.92                & 1.54          & 0.07              & 3.72                \\
9                                                           & \{acknowledgement, choice\}          & \{objective\} & 0.92                & 1.54          & 0.07              & 3.72                \\
10                                                          & \{acknowledgement, data\}            & \{objective\} & 0.91                & 1.52          & 0.08              & 3.54                \\
11                                                          & \{puzzles, acknowledgement\}         & \{objective\} & 0.9                 & 1.51          & 0.08              & 3.38                \\ \hline
\end{tabular}
\end{adjustbox}
\label{tab:f}
\end{table*}

As for the male sample, Objective was also the main element but, in contrast to the females, we did not find any rules (with confidence $>$ 90\%)  related to  elements that were most relevant to the male population (Table \ref{tab:j}). There was only 1 rule that specified a social element (Social Pressure) amongst all the 14 rules. We can observe that Progress appears in almost all the rules, followed by Choice, appearing in seven rules. This means that, in our sample, designers and teachers should consider quests and missions that contain a form of progression and allow the students to make meaningful choices; those choices can be tied to a challenge (Rule 14), to transactions (Rule 24) and points (Rule 16).

\begin{table*}[!ht]
\caption{Relevant rules to male sample}
\centering
\begin{adjustbox}{max width=\textwidth}
\begin{tabular}{lllllll}
\hline
\textbf{\begin{tabular}[c]{@{}l@{}}Rule \\ ID\end{tabular}} & \multicolumn{1}{c}{\textbf{If}}                & \textbf{Then} & \textbf{Conf} & \textbf{Lift} & \textbf{Lev} & \textbf{Conv} \\ \hline
12                                                          & \{renovation, progress, choice\}               & \{objective\} & 0.94          & 1.6           & 0.04         & 5.37          \\
13                                                          & \{progress, social pressure, data\}            & \{objective\} & 0.93          & 1.59          & 0.04         & 5.04          \\
14                                                          & \{progress, puzzles, acknowledgement\}         & \{objective\} & 0.93          & 1.59          & 0.05         & 5.16          \\
15                                                          & \{level, renovation, progress\}                & \{objective\} & 0.93          & 1.59          & 0.04         & 4.96          \\
16                                                          & \{point, objective, puzzles\}                  & \{level\}     & 0.92          & 1.93          & 0.05         & 5.74          \\
17                                                          & \{level, progress, puzzles, choice\}           & \{objective\} & 0.92          & 1.57          & 0.04         & 4.27          \\
18                                                          & \{progress, acknowledgement, data\}            & \{objective\} & 0.91          & 1.55          & 0.05         & 4.13          \\
19                                                          & \{point, progress, choice\}                    & \{objective\} & 0.91          & 1.55          & 0.04         & 4.01          \\
20                                                          & \{progress, novelty, data, choice\}            & \{objective\} & 0.91          & 1.55          & 0.04         & 4.01          \\
21                                                          & \{progress, novelty, acknowledgement, choice\} & \{objective\} & 0.91          & 1.55          & 0.04         & 3.95          \\
22                                                          & \{renovation, progress, novelty\}              & \{objective\} & 0.91          & 1.55          & 0.04         & 3.92          \\
23                                                          & \{level, progress, data, choice\}              & \{objective\} & 0.91          & 1.55          & 0.04         & 3.92          \\
24                                                          & \{progress, novelty, economy\}                 & \{objective\} & 0.91          & 1.54          & 0.04         & 3.78          \\
25                                                          & \{progress, choice, economy\}                  & \{objective\} & 0.91          & 1.54          & 0.04         & 3.78          \\ \hline
\end{tabular}
\end{adjustbox}
\label{tab:j}
\end{table*}

Based on the data on Tables \ref{tab:f} and \ref{tab:j}, we can observe that Objective associated to Progress is a concept that is (generally) well accepted by both genders. This means that teachers and designers should focus on, \textit{e.g.}, developing quests (which can be tied to their original learning objectives) that allow the learners to place themselves within the task. This is important, since in some educational context, students do not know why they are learning a specific content; and consequently,  may become demotivated \cite{Hardre2012}. In practice, this means that teachers can create milestones or goals, allowing students to visualise their progress towards this goal.
Thus, guidelines can be provided to  teachers,  to convert their objectives in their classes into milestones or quests. Additionally,  other representations of Progress,  showing the users where they are in the course could be implemented, such as those supported by Levels, Points and Data. %% We can observe that some of those rules were applied by Toda \textit{et al}. \cite{Toda2018f} where the authors used elements like Level, Progress and Puzzles to achieve specific Objectives in their approach and achieved positive results regarding the acceptance by the students. However they also tied their gamification instance to Competition, which hindered some students' satisfaction with the proposed gamification. It is worth to mention that this Competition element was not found amongst any of our 25 rules.

\subsection{Further Discussion}
\label{sec:dis}
We consider this work to be important, as, with the advent of 'big data', various theoretical assumptions and statements can now be backed up by (significant) evidence. In the case of game elements, there is firstly a vast (not always research-based) evidence that games are linked to motivation, and keep players 'in the flow' \cite{Mihaly}. %% add Czichenmickali reference
Some studies even link specific game elements to higher levels of commitment or motivation \cite{Yee2006}. %% add reference
Based on this evidence, as well as theories of motivation, gamification has been proposed for education. Currently, however, the data supporting these assumptions is scarce. There is a lot of small-scale empirical evidence, at classroom-scale, of approaches that showed mixed successes \cite{DeSousaBorges2014,Dichev2017a}. %% add reference
In a similar way, there is evidence that gamification can also have undesirable effects \cite{Toda2018c}. This clearly points to the fact that there are parameters which need taken into consideration, which may influence the outcomes of gamified approaches to education. In this study, we specifically focus on demographic parameters - namely, gender. 

Gender in education has been brought to the fore recently, with the advent of initiatives such as the 'Athena SWAN'\footnote{https://www.ecu.ac.uk/equality-charters/athena-swan/} initiative towards gender equality in Higher Education in the UK, as well as similar initiatives world-wide. Importantly, equality doesn't mean 'one size fits all': on the contrary, gender equality means that the provision of education takes into account specific preferences that may be gender related. %\textit{E.g}., giving only football -and beer- related examples when learning programming may appeal to a certain demographic and not another. Interspersing such examples with fashion and romance-movie-related examples may widen the demographics-related appeal, etc. 
In a similar vein, certain types of games appeal to certain demographics and not others. For instance, card-related games are potentially more appealing to women, and first-player-shooter games to men (although, of course, preferences can vary) \cite{esa2018}. %% do we have any references to this?

\begin{comment}
In terms of the study performed, it may be interesting to further analyse the surveyees who didn't specify their age (by using out of range answers), especially the ones who identified themselves as 'female' - as women often don't like to mention their age, and thus there might be more respondents that are women 'hidden' amongst those values. It may be interesting to find out if their preferences coincide with the group that did specify their age.

Further-on, we could reanalyse users with ages above 90, to find out if we can extract some interesting specific preferences for this group (and thus, if there really were users that had that age, or just erroneous inputs). In general, age demographics in terms of preferences for game elements is interesting in itself, and is more relevant now in the age of Massive Online Open Courses (MOOCs), where online learning often is life-long-learning. 
\end{comment}

Further analysis of Table \ref{tab:b} shows that male and female preferences of some elements is relatively similar; e.g., Data, Economy, Objective are almost identical, and some are only slightly different. Thus, some game elements may be perceived  similarly by males and females - which makes the teacher's job much easier, in terms of design choices. This also puts more emphasis on the game elements where large differences exist, as well as on game elements where the differences in preference are slight, but statistically relevant.

Some of the results obtained were surprising: for instance, we expected females to appreciate cooperation more than males, but results (see section \ref{sub:des}) showed otherwise. We did, on the other hand, obtain the expected results in terms of preference for competition. It is possible that online social interaction overall is perceived differently by males and females; for instance, females may perceive any type of social interaction online, where people are not known in advance, and anyone from anywhere can participate, as potentially threatening. These types of areas need further analysis. 

For educational applications, it may seem that such potential 'fears' are less likely in controlled (classroom, or classroom-based) environments. However, for example, on Massive Online Open Courses (MOOCs), where people can participate from anywhere, such issues can again prevail. In fact, research on social interactions on MOOCs (e.g., comments, etc.) shows a predominance of males performing such activities. %% Please find a reference!
In contrast, females preferred puzzles (which can be solved also as solo-player) and the 'Renovation' element (see Figure \ref{fig:c}), which allow for an independent style of play where one focuses only on ones own progress, instead of being interrupted by others.

\section{Conclusions}
\label{sec:con}
This work presents our approach based on DDGD, towards planning of gamification in the educational information systems domain by using data mining.  The main contribution of our work is to present a poll of gamified strategies tied to male and female genders.
%to show, based on a large-scale study (larger than the recommended sample size),  that when planning gamification strategies, it is necessary to take into consideration user differences in terms of  gender, when selecting specific game elements. Moreover, this is the first data-driven study of game elements in perception applied in education domain, to the best of our knowledge. We have shown that there are statistically significant differences in preferences of these game elements between men and women, and that specific, dissimilar association rules emerge for each gender, based on the game elements. 
Furthermore, we use real data to aid in the decision process of teachers and instructors is selecting gamification strategies. Through our data, we could identify that males would make more use of social interactions, with strong confidence rules pairing  gamification elements Progression and Choice.  For the females, we identified that user experience and rewards are more relevant, with association rules indicating a strong confidence for the need of Acknowledgement and Progression. We believe that this work can impact the way teachers perceive and apply gamification in their environments, consequently improving students' engagement and motivation through a game-like experience.

%Therefore, we believe this is an important initial step towards better understanding relations between game elements perception, based on user characteristics, and exploring how they may impact on users’ behaviour and acceptance. Moreover, our analysis is aimed at aiding teachers and designers in adopting gamification strategies within their learning contexts (e.g., classrooms or educational systems), based on the gender of their learners. The dataset, tables and statistics that were used in this work can be found at [link omitted due to Blind Review].  

%% For future work, we plan to create scenarios to validate these rules and verify how they impact on users' behaviour, motivation, engagement and performance. We are also developing a recommendation model based on these rules and trigger them in an automatic way for teachers and designers, who then would not need to study or spend time understanding the concepts beneath gamification. We also aim at collecting more data, focusing on other genders to expand the recommendations. 

\section*{Acknowledgement}
We would like to thank FAPESP (Projects 2016/02765-2; 2018/11180-3; 2018/07688-1) for the funding provided.

\bibliographystyle{IEEEtran}

\bibliography{references}

\end{document}